\documentclass[aps,showpacs]{revtex4}

\usepackage{amssymb}
\usepackage{graphicx}
\usepackage{dcolumn}
\usepackage{bm}
\usepackage{amsmath}
\usepackage{color}
\usepackage{color}
\usepackage{subfigure}

\def\br{{\bf r}}
\def\bq{{\bf q}}

\begin{document}

\title{Scaling behavior of disordered lattice fermions in two dimensions}
 
 \author{Antonio Hill}
 \author{Klaus Ziegler}

 \affiliation{Institut f\"ur Physik, Universit\"at Augsburg, D-86135 Augsburg, Germany}
 \date{\today }
\begin{abstract}
We propose a lattice model for Dirac fermions which allows us to break the degeneracy of the node structure. 
In the presence of a random gap we analyze the scaling behavior of the
localization length as a function of the system width within a numerical transfer-matrix approach. 
Depending on the strength of the randomness, there are different scaling regimes for weak, intermediate and 
strong disorder. These regimes are separated by transitions that are characterized by one-parameter scaling. 
\end{abstract}
\maketitle





\section{Introduction}

Two-dimensional Dirac fermions play a crucial role in graphene and on the surface of topological insulators.
A fascinating observation in graphene is the robust electronic transport in the vicinity
of the two Dirac nodes, where two electronic bands meet each other with linear dispersion 
\cite{novoselov05,zhang05}.
The latter is a consequence of the honeycomb lattice in graphene, which decomposes into two triangular lattices.

In contrast to the experimentally observed robust transport properties 
it has been claimed from the theoretical side that transport is very sensitive whether inter-node
scattering is present or not in the presence of disorder \cite{ando98}. In particular, there has been speculations that
electronic states are delocalized in the absence of inter-node scattering but localized in its presence.
This has been explained by changing the symmetry class of the underlying Hamiltonian from symplectic to orthogonal 
\cite{ando02,beenakker08}.
These claims are based on weak-localization calculations \cite{ando98,ando02}, which
predict weak (anti-) localization (with) without inter-node scattering. Since weak localization
calculations can only indicate the tendency towards localization, it would be interesting to
evaluate this effect directly in terms of the scaling behavior of the localization length.  
For this purpose we shall study the localization
length of a strip of finite width $M$ under a change of $M$ in this paper. Our method, originally introduced
for transfer-matrix calculations of the Schr\"odinger Hamiltonian \cite{Pichard1981, MacKinnon1983}, will 
be applied subsequently to 2D lattice Dirac fermions with one or more nodes.
For this purpose we introduce a model which has two bands and four Dirac nodes.
We can open a gap at one node and gaps for the other three nods independently. This allows us to study the
effect of intervally scattering by either keeping all four nodes or removing three of them and keeping only
a single node. 

The aim of this work is to understand the scaling behavior of the localization length in two 
dimensions in the metallic regime and near a metal-insulator transition due to a gap opening. 
The latter has been observed recently in graphene 
\cite{elias2009,bostwick09,geim11}. where it appears in the presence of a random gap in  
the Dirac spectrum. If the average gap value is small in comparison to the fluctuation strength
the system is metallic whereas it is insulating when the gap fluctuations are too weak in comparison
to the average gap \cite{Ziegler2009,abergel10}.

\section{Model}

A tight-binding description of electrons in graphene yields the famous energy dispersion with two 
separate nodes (or neutrality points) in the Brillouin zone. In the vicinity of these nodes the
momentum dependence of the spectrum is found to be linear and the low--energy behavior 
of quasi particles can well be described by the Dirac equation
$H \psi(x,y) = E \psi(x,y)$ 
with the Hamiltonian
\begin{equation}
H=  -i\hbar v_F \,(\vec{\sigma} \cdot \vec{\nabla)} + v_F^2\, m \, \sigma_3\, ,
\label{eq:plain-hamiltonian}
\end{equation}
where $v_F$ is the Fermi velocity, $\vec{\sigma}$ is the vector of Pauli matrices and $\psi=(\varphi_1, \varphi_2)$ 
is the two component spinor wave function, furthermore we set $\hbar v_F=1$. 

A numerical treatment of the Dirac equation requires a discretization in space.
However, the naive discretization through replacing the differential operator by a difference operator
leads to additional new nodes, which is often called fermion doubling or multiplication ~\cite{Susskind1977}. 
In real space there are two methods to circumvent this problem~\cite{Stacey1982,beenakker08,beenakker10}. One that we will describe in this section goes back to the idea of Susskind.
We start with discretizing the differential operator in a symmetric way
\begin{equation}
 \partial_x f(x)\approx \frac{1}{2\Delta} (f_{l+\Delta} - f_{l-\Delta} )
 \ ,
\end{equation}
where $\Delta$ is the lattice constant which we set to unity in the following. The discrete Dirac equation for $m=0$
then takes the form
\begin{align*}
 -\frac{i}{2} \sigma_1 \left\lbrace \psi_{l+1,n} - \psi_{l-1,n}  \right\rbrace -\frac{i}{2} \sigma_2 
\left\lbrace\psi_{l,n+1} - \psi_{l,n-1}  \right\rbrace = E\sigma_0\psi_{l,n} \, 
\end{align*}
with lattice points given by the coordinates $(l,n)$ with integer $l$ and $n$. Fourier transformation leads to eigenvalues 
$E=\pm \sqrt{sin(k_x)^2 + sin(k_y)^2}$ which have four Dirac cones in the Brillouin zone corresponding to four Dirac fermions. 
In order to open a gap at three of them we introduce a lattice operator~\cite{Ziegler1996} which acts on a wave function as
\begin{equation}
 \hat{B} \, \psi_{l,n} = \frac{1}{2}\left\lbrace \psi_{l+1,n} + \psi_{l-1,n} + \psi_{l,n+1} + \psi_{l,n-1}  \right\rbrace \, .
\end{equation}
Now we discretize Hamiltonian~(\ref{eq:plain-hamiltonian}) by including the lattice operator $\hat{B}$ and a random gap term
\begin{equation}
 H \rightarrow H + \delta(\hat{B} - 2)\sigma_3 + m_{l,n}\,\sigma_3 \, .
\end{equation}
For uniform gap $m$ our new Hamiltonian reads in Fourier representation
\begin{equation}
  H= \begin{pmatrix}
     m+\delta (cos(k_x)+cos(k_y)-2)  & \sin(k_x) + i\sin(k_y) \\  \sin(k_x) - i\sin(k_y) & -m-\delta (cos(k_x)+cos(k_y)-2)
    \end{pmatrix}
    \label{main_ham}
\end{equation}
with the dispersion
\begin{equation}
 E = \pm \sqrt{sin(k_x)^2 + sin(k_y)^2 + (m+\delta cos(k_x)+\delta cos(k_y)-2\delta )^2} \, .
\label{eq:dispersion}                                                                                            
\end{equation}
For $m=0, \delta\ne0$ there is a node at $k_x=k_y=0$ and three additional nodes for $m=0, \delta=0$ at $k_x,k_y=\pm\pi$ (cf. Fig.).
Using this model node degeneracy can be lifted via the parameter $\delta$.

We absorb the index $n$ with the help of matrix representation and write for the wave function
\begin{equation}
  \psi_{l+1} = H^{Y} \ \psi_l + H^{D} \ \psi_{l-1} \, .
\end{equation}
Each spinor component is now a $M$-component vector, where $M$ is the width of a strip and thus $n=1,2,...,M$. The matrices $H^{Y}$, $H^{D}$ read
\begin{align*}
 H^{Y}_{n,n} &= 2 S^{-1}\left[  E\, \sigma_0  + (2\delta-m)\sigma_3 \right] &&  H^{Y}_{n,n+1} = S^{-1} \left[i\sigma_2 -\delta \sigma_3  \right] \\
 H^{Y}_{n,n-1} &= -  S^{-1}\left[ i\sigma_2 + \delta\sigma_3  \right]  && H^D_{n,n} = - S^{-1}\left[i \sigma_1  + \delta \sigma_3  \right] \, 
\end{align*}
with $S=-i\sigma_1 + \delta \sigma_3$ and where $H^{Y}$ has periodic boundary conditions in the $y$-direction.
This matrix structure allows us to construct a transfer matrix $T_l$ through the equation \cite{MacKinnon1983}
\begin{equation}
 \begin{pmatrix}\psi_{l+1} \\ \psi_{l} \end{pmatrix} = \begin{pmatrix}H^Y & H^D \\ 1 & 0 \end{pmatrix}
 \begin{pmatrix}\psi_l \\ \psi_{l-1} \end{pmatrix} \equiv T_l \begin{pmatrix}\psi_l \\ \psi_{l-1} \end{pmatrix}\, .
\label{eq:transfer-matrix-2d}
\end{equation}
The introduction of a different random potential, e.g. random scalar potential, is straight forward.

\subsection{Lyapunov exponents}

According to \cite{Pichard1981, MacKinnon1983} the transfer matrices $T_l$, defined in 
Eq.~(\ref{eq:transfer-matrix-2d}), can be used to calculate Lyapunov characteristic exponents (LCE). With initial values 
$\psi_0$ and $\psi_1$ the iteration of Eq.~(\ref{eq:transfer-matrix-2d}) provides $\psi_L$ by the product matrix 
\begin{equation}
 M_L = \prod_{l=1}^{L} T_l \, .
\end{equation}
For disordered systems this is a product of random matrices that satisfies Oseledec's theorem \cite{oseledec}. The latter states
that there exists a limiting matrix 
\begin{equation}
\Gamma= \lim_{L \rightarrow \infty} (M_L^{\dagger} M_L)^{1/2L}  \, . 
\label{eq:oseledec}
\end{equation}
The eigenvalues of $\Gamma$ are usually written as exponential functions $\exp(\gamma_i)$, where $\gamma_i$ is the LCE. Adapting the numerical algorithm described in 
\cite{MacKinnon1983}, the whole Lyapunov spectrum can be calculated and the smallest LCE is identified with the inverse localization length \cite{Pichard1981}. 

\begin{figure}[ht]
\centering
 \subfigure{
   \includegraphics[scale=0.65]{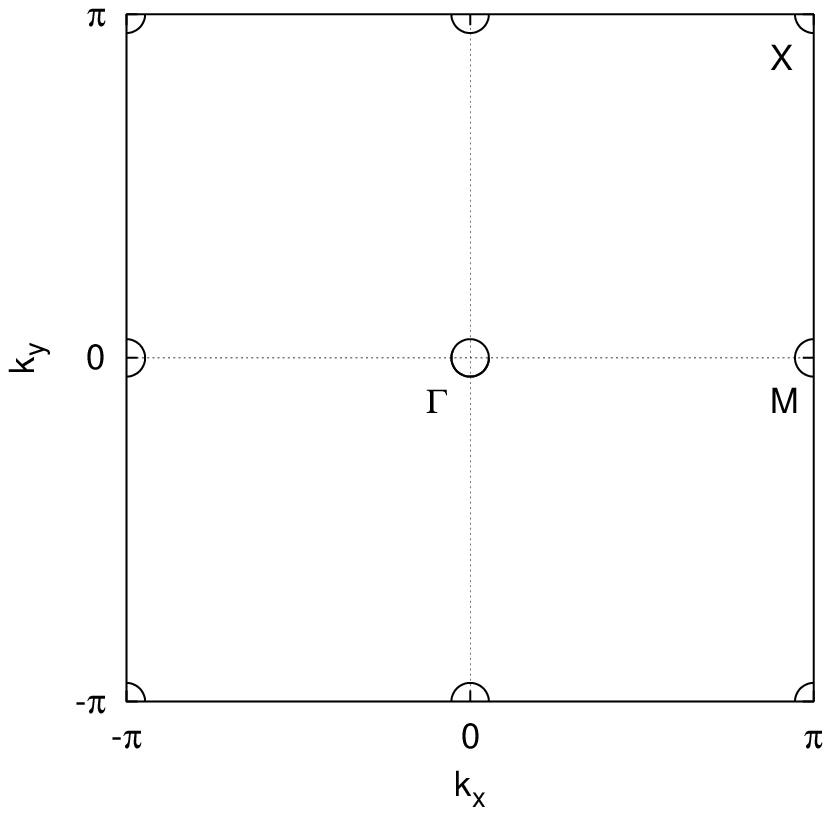}
 }
\subfigure{
 \includegraphics[scale=0.65]{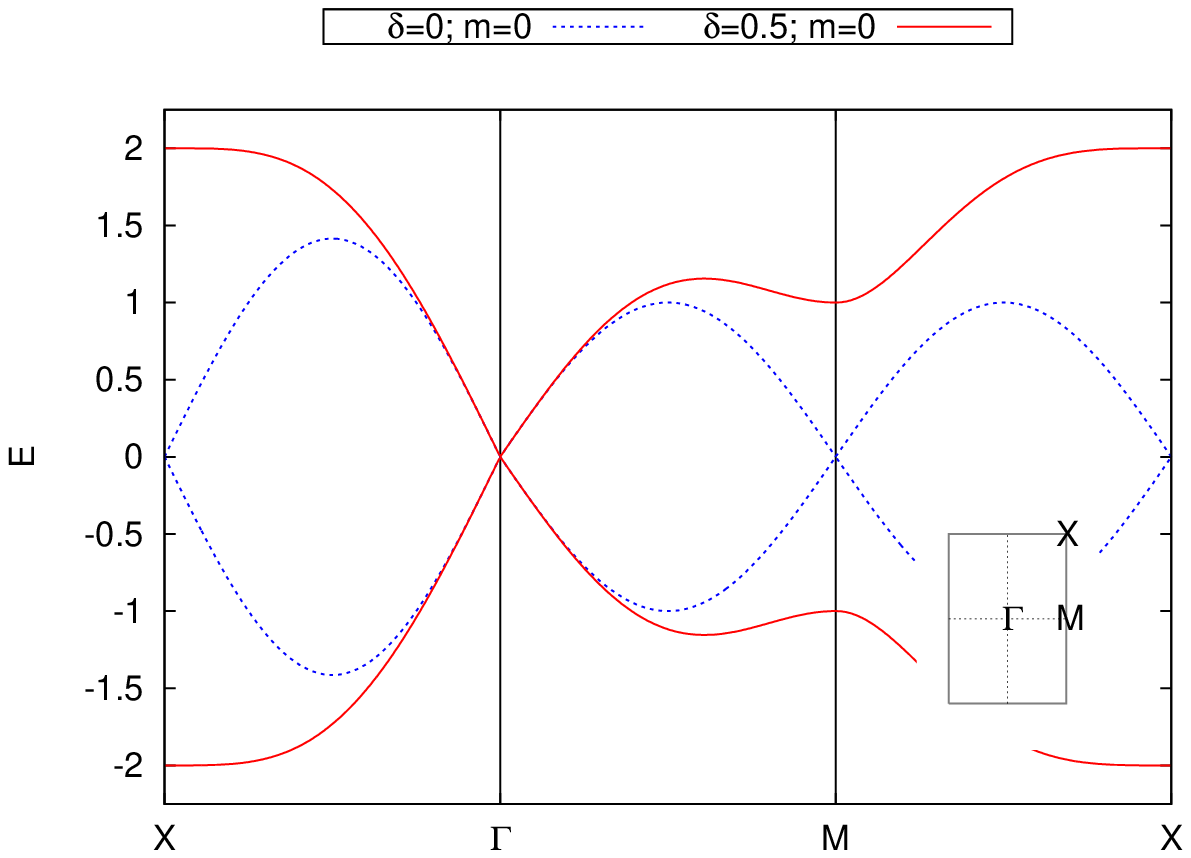}
 }
\caption{Brillouin zone of the discrete Dirac equation with circles depicting the positions of the Dirac cones (left). 
A cut through the energy dispersion~(\ref{eq:dispersion}) is shown on the right for $\delta=0$ (blue line), 
$\delta=0.5$ (red line). 
}
\label{fig:dispersion-vergleich}
\end{figure}

\section{Numerical results for random gap}

After introducing the model and the corresponding transfer matrices we calculate the inverse of the smallest 
LCE $\Lambda=1/\gamma_{min}$ 
which is identified as the localization length. 
$\Lambda$ increases with the system width $M$ according to a power law:
\begin{equation}
 \Lambda \propto M^{\alpha} \, ,
\label{eq:LL-power-law}
\end{equation}
where $\alpha>1$ ($\alpha<1$) in the regime of extended (localized) states, and $\alpha=1$ in the critical regime.
For the exponentially localized regime we expect $\Lambda \propto const$. 
According to the one-parameter scaling theory by MacKinnon~\cite{MacKinnon1981},
the normalized localization length $ \tilde{\Lambda}=\Lambda/M$ obeys the equation
\begin{equation}
 \frac{d\ln \tilde{\Lambda}}{d\ln M} = \chi(\ln \tilde{\Lambda})
 \ ,
\label{scaling-eq}
\end{equation}
where $\chi$ is an unknown function with solutions of the form
\begin{equation}
 \tilde{\Lambda}(M,W) = f(\xi(W)/M) \, .
\label{eq:scaling-sol}
\end{equation}
The parameter $W$ characterizes the disorder strength and $\xi$ is a characteristic length of the system. 
The one-parameter scaling theory states that $\tilde{\Lambda}$ is not depending 
on $M$ and $W$ separately. Any change of disorder strength $W$ can be compensated by a change of the system width $M$. 
Furthermore, from the behavior of $\tilde{\Lambda}$ in the vicinity of a scale-invariant point it is possible to calculate the 
critical exponent $\nu$ of the correlation length~\cite{MacKinnon1983}, which is the localization length of the infinite system. 
This is done by Taylor expansion
\begin{align}
 \ln  \tilde{\Lambda} &= \ln  \tilde{\Lambda}_c + \sum_{s=1}^{S} A_s \left(|W-W_c|M^{1/\nu}\right)^s \\
    & =  \ln  \tilde{\Lambda}_c + \sum_{s=1}^{S} A_s \left(\frac{\xi}{M}\right)^{-s/\nu}\, ,
\label{eq:fit-exponent}
\end{align}
with $\xi = |W-W_c|^{-\nu}$. 
Comparing the latter with eq.~(\ref{eq:scaling-sol}), the scaling function $\xi$ can be interpreted as the characteristic length scale.


\subsection{Preserved node symmetry: $\delta=0$}

In this case we have a four-fold degeneracy of the node structure.
First we calculate $\Lambda$ from transfer matrix~(\ref{eq:transfer-matrix-2d}) with $\delta=0$. 
If it is not mentioned explicitly we use for the random gap $m$ a box distribution on the interval $[{\bar m}-W/2,{\bar m}+W/2]$, 
where the corresponding variance is given by $W^2/12$. Furthermore, we restrict our 
calculations to the Dirac point (i.e. $E=0$). 

\begin{figure}[ht]
\centering
\subfigure{
 \includegraphics[scale=0.6]{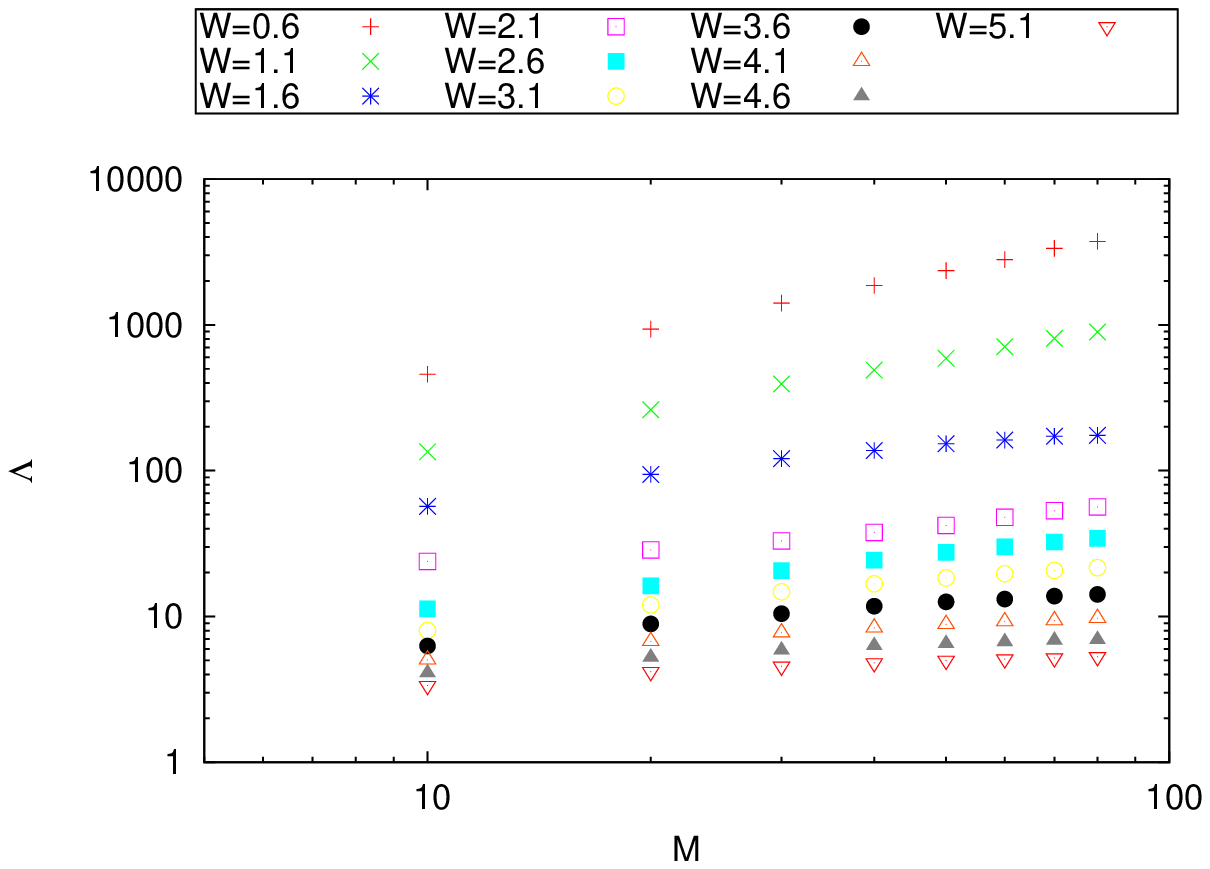}
\label{fig:ll-gap0-delta0-box}
 }
 \subfigure{
 \includegraphics[scale=0.6]{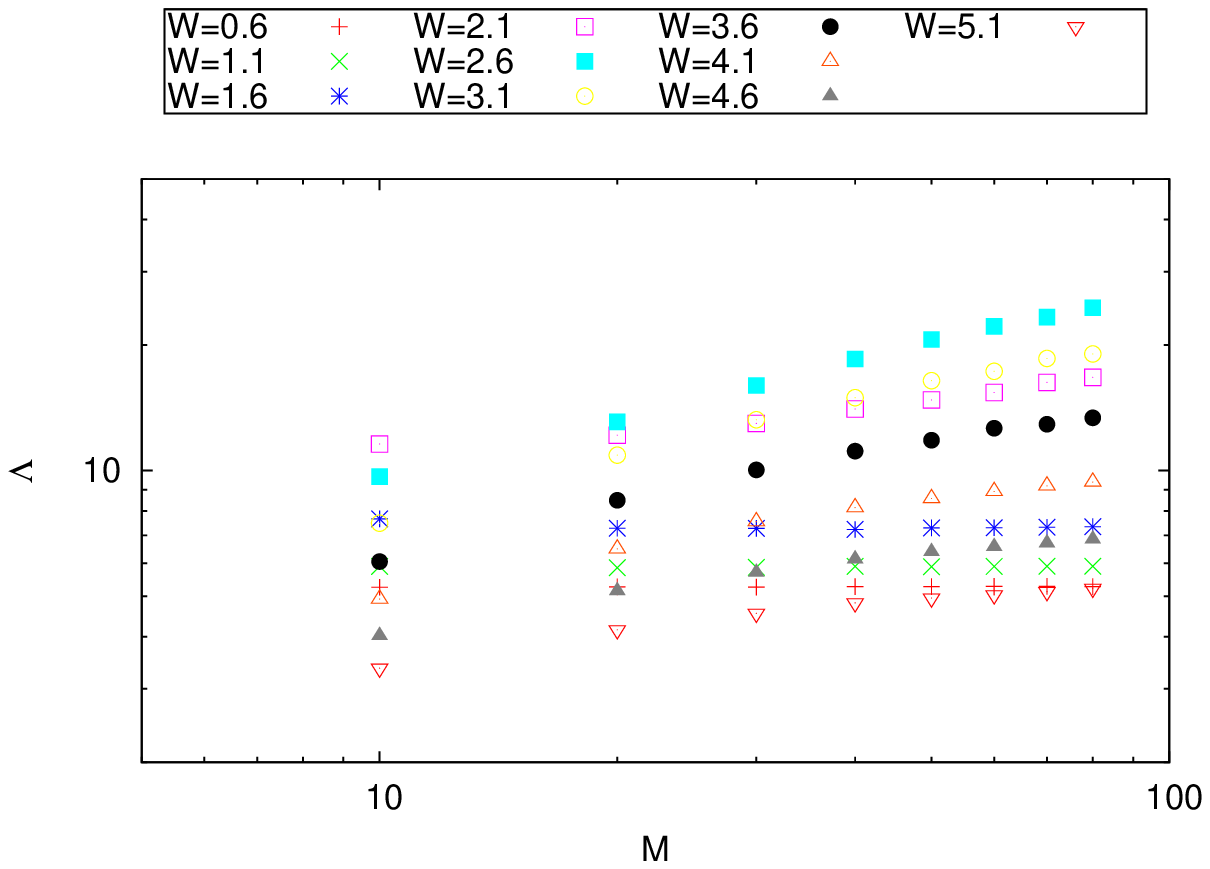}
\label{fig:rescaled-nll-gap0-delta0}
 }
\caption{Localization length for preserved node degeneracy ($\delta=0$) with average gap $\bar{m}=0$ (left panel) and $\bar{m}=0.2$ (right panel)
as a function of strip width $M$. 
}
\label{fig:scaling-dirac-gap0-delta0a}
\end{figure}

\begin{figure}[ht]
\centering
\subfigure{
 \includegraphics[scale=0.6]{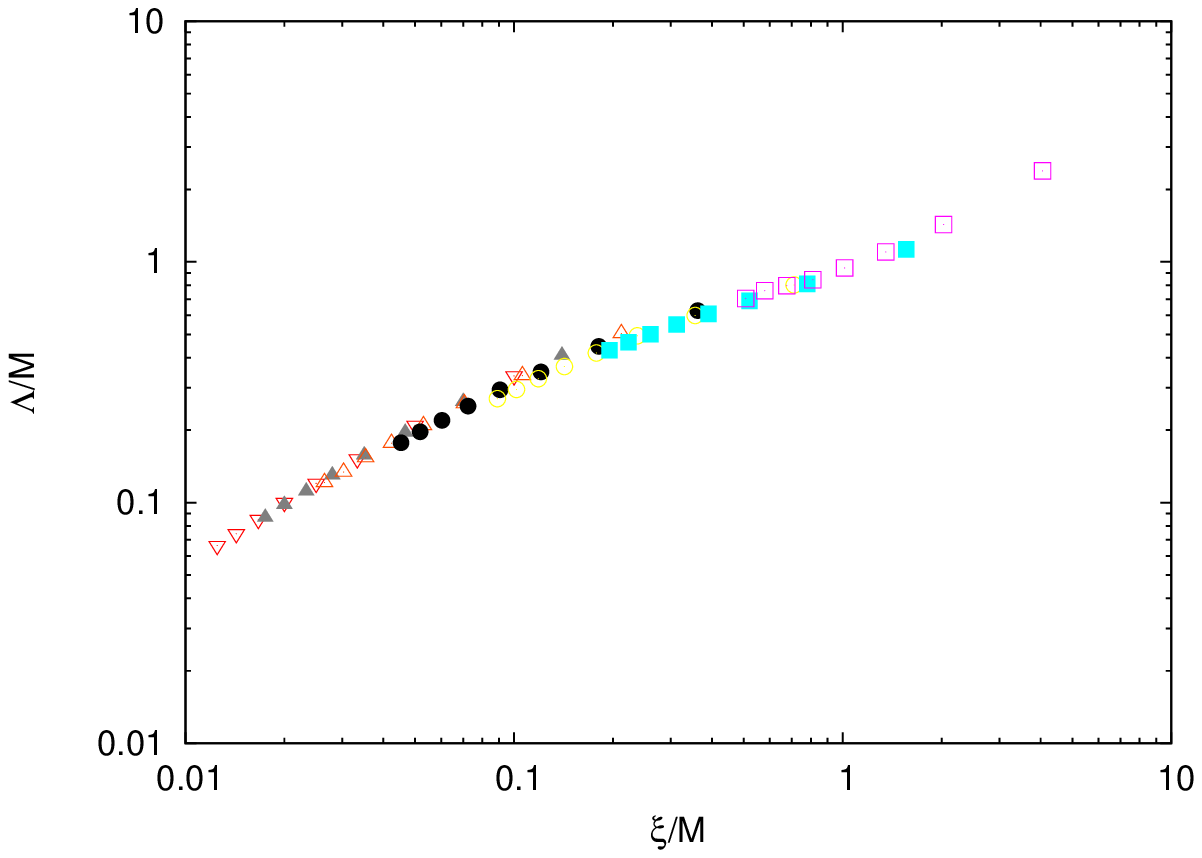}
\label{fig:rescaled-nll-gap0-delta0a}
 }
 \subfigure{
  \includegraphics[scale=0.6]{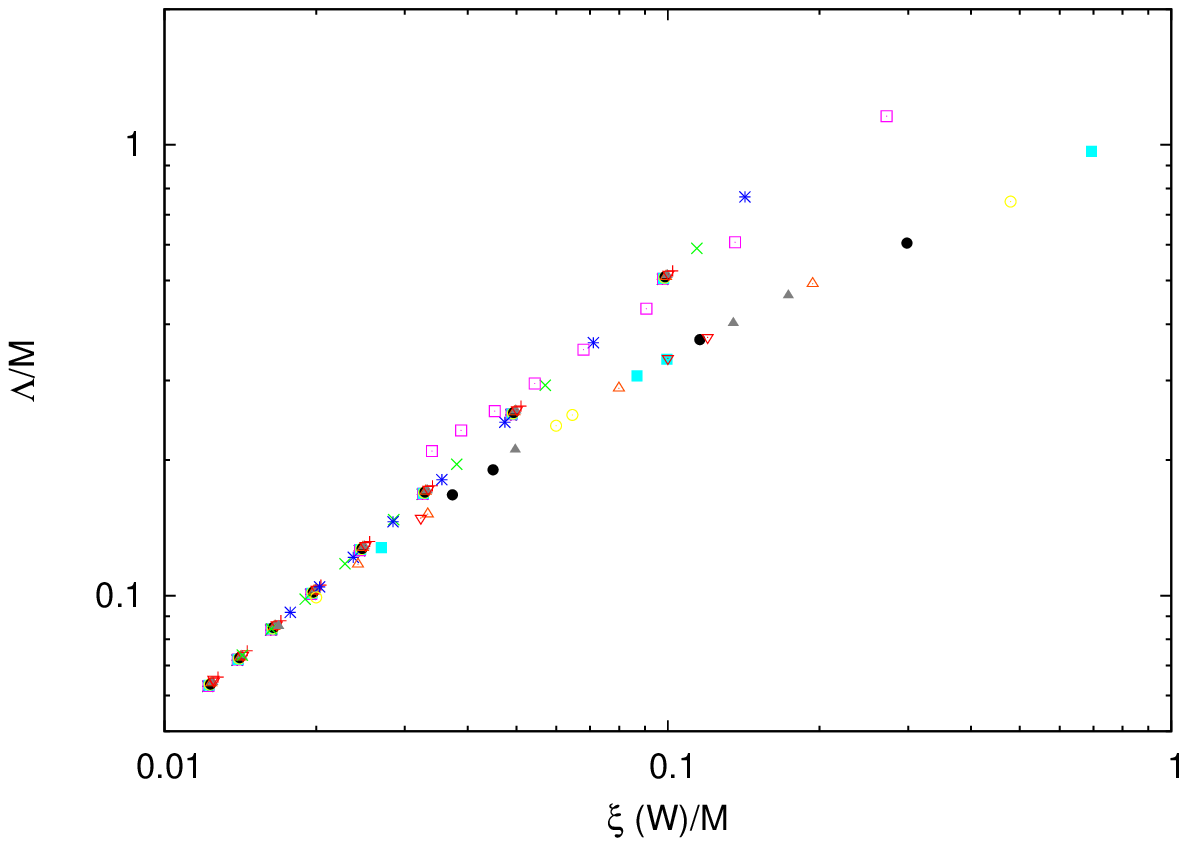}
\label{fig:rescaled-data-dirac-gap02-delta0_a}
}
\caption{Scaling plot of the localization length for $\delta=0$, $\bar{m}=0$ (left) and $\delta=0$, $\bar{m}=0.2$ (right).
Left: Rescaled without data for $W=0.6; 1.1; 1.6$.}
\label{fig:scaling-dirac-gap0-delta0}
\end{figure}

\begin{figure}[ht]
\centering
 \subfigure{
    \includegraphics[scale=0.65]{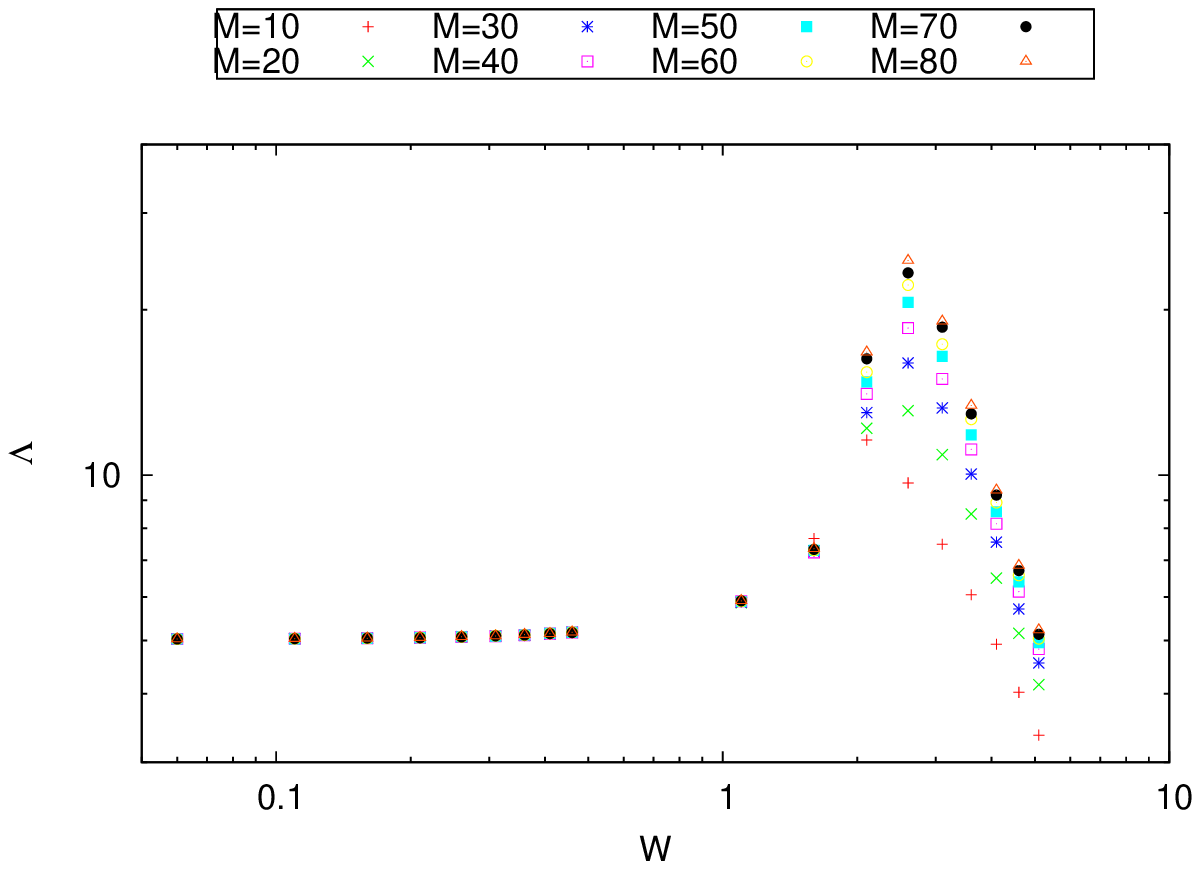}
  \label{fig:nll-gap0-delta05-raw_a}
}
  \subfigure{
  \includegraphics[scale=0.65]{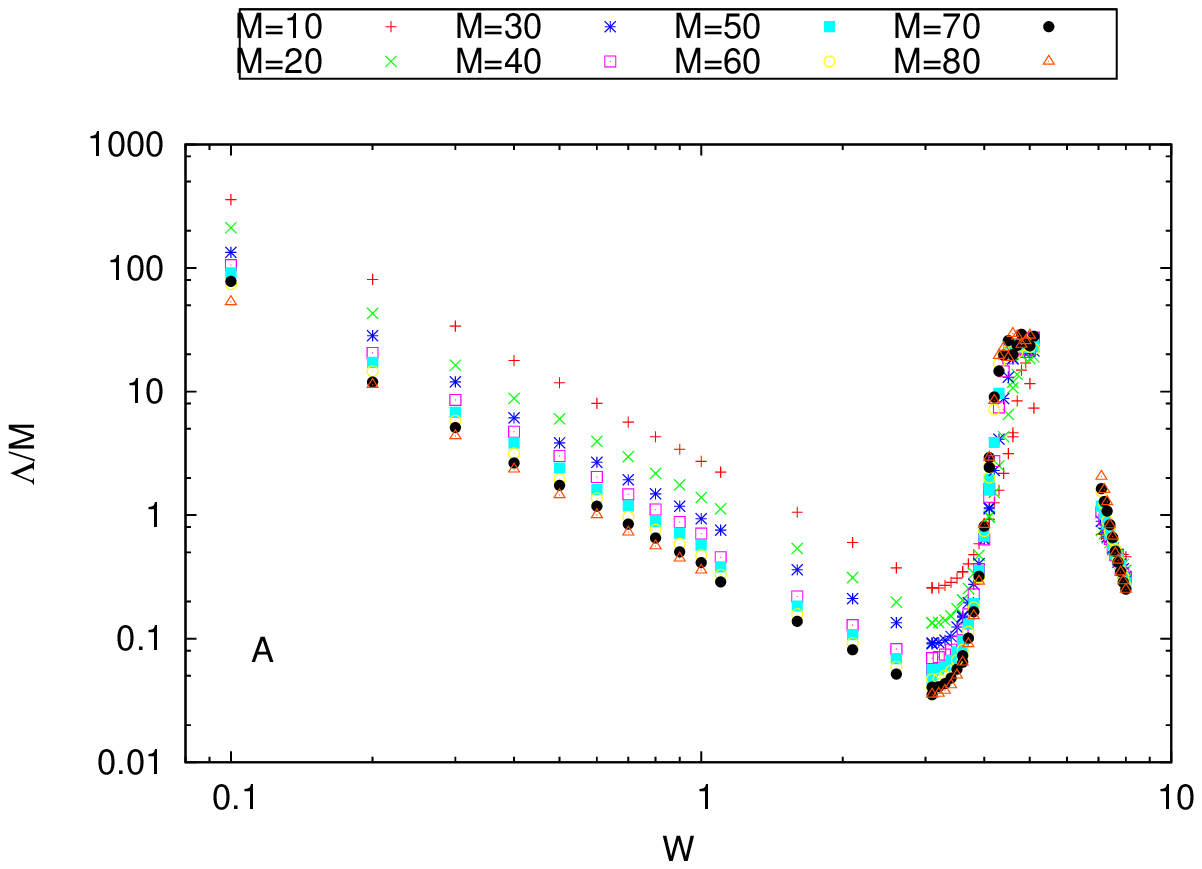}
  \label{fig:loc-len-over-disorder-gap0-delta05_a}
}
    \subfigure{
  \includegraphics[scale=0.65]{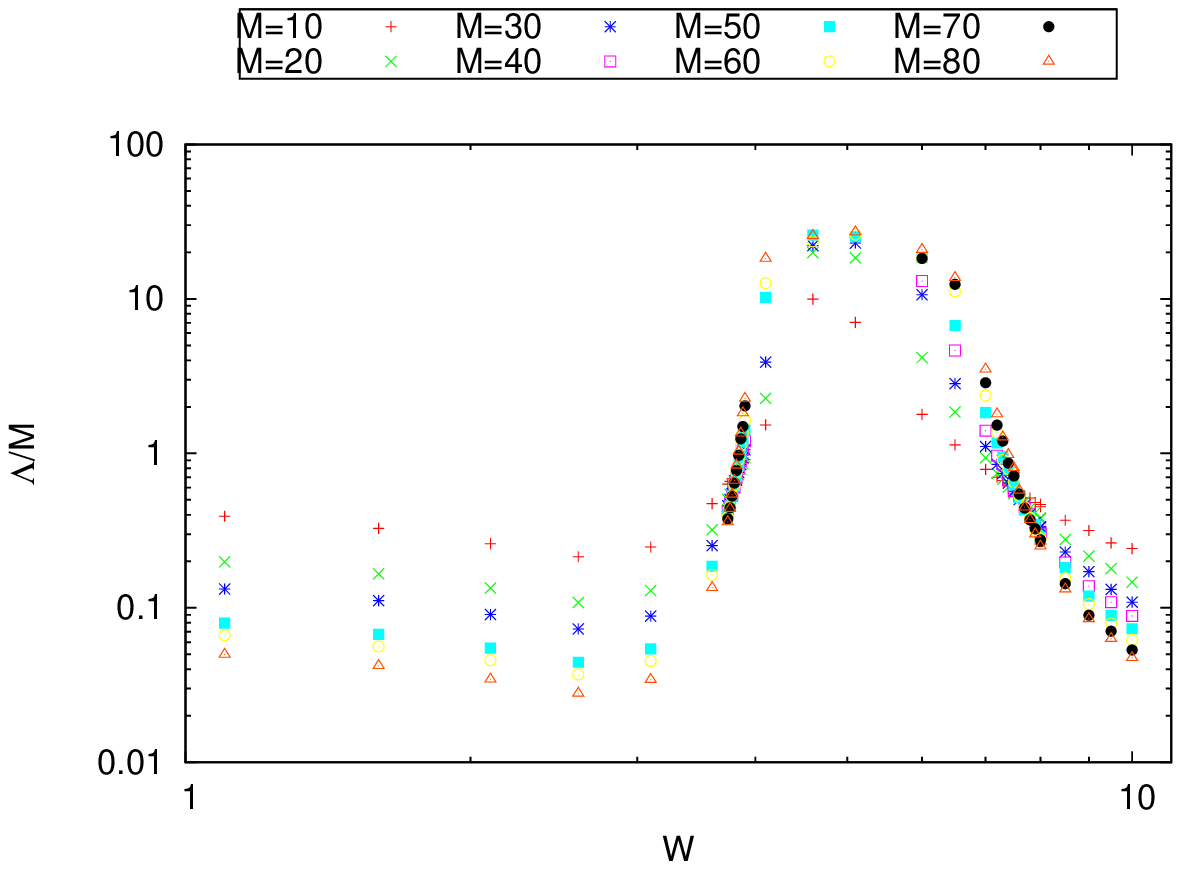}
  \label{fig:loc-len-over-disorder-gap02-delta05_a}
}
\caption{Localization length for random gap with zero mean and broken node symmetry with ($\delta=0.0$, $\bar{m}=0.2$) (left),
 $\delta=0.5$ and $\bar{m}=0$ (right) and  $\delta=0.5$ and $\bar{m}=0.2$ (bottom) as a function of disorder.}
\end{figure}

Fig.~\ref{fig:scaling-dirac-gap0-delta0a} depicts the effect of the average gap ${\bar m}$ on the localization length $\Lambda$.
The localization length always increases with system width $M$, indicating that there is no exponential localization.
Only for very weak disorder ($W<0.2$) and ${\bar m}=0.2$ the localization length $\Lambda$ is almost independent of
$M$, which indicates
exponential localization for ${\bar m}=0.2$ (cf. Fig.~\ref{fig:nll-gap0-delta05-raw_a}).
As disorder increases the localization length decreases monotonically for ${\bar m}=0$ but not for ${\bar m}=0.2$
(cf. Fig.~\ref{fig:rescaled-nll-gap0-delta0}. 
If we normalize $\Lambda$ by strip 
width $M$ and perform single parameter scaling as described in~\cite{MacKinnon1983}, 
almost all data points collapse to a single curve (cf. Fig.~\ref{fig:rescaled-nll-gap0-delta0a}). However,
we had to neglect data points from weak disorder ($W\leq1.6$) to see clearly a scaling behavior. 

The behavior of $\Lambda$ for a nonzero average gap ($\bar{m}=0.2$) is more complex, as shown in 
Figs.~\ref{fig:rescaled-data-dirac-gap02-delta0_a}, \ref{fig:nll-gap0-delta05-raw_a}. 
For weak disorder the localization length converges to a 
constant value for increasing $M$. As disorder increases $\Lambda$ increases also but remains constant for large $M$. 
Then there is a transition at $W\approx 2.1$ where $\Lambda$ is again growing with system size but the slope decreases 
with increasing disorder. 

Due to this behavior of $\Lambda$ as a function of disorder it is not possible to perform single parameter scaling 
in the common way. One approach to calculate the scaling function is to minimize the variance of $\ln M - \ln \xi$ 
for each localization length~\cite{MacKinnon1981}. In a double logarithmic plot of $\tilde{\Lambda}$ the problem 
of one parameter scaling translates then into shifting all curves onto one~\cite{MacKinnon1983}. Since the position 
of the resulting curve is irrelevant it is convenient to shift all curves onto the lowest i.e. that for biggest disorder. 
If one looks closely at the data in Fig.~\ref{fig:rescaled-nll-gap0-delta0} one sees that this is not possible only by shifting. 
Comparing to Fig.~\ref{fig:nll-gap0-delta05-raw_a} one can distinguish two regimes separated at $W\approx 2.1$. 
In both regimes one parameter scaling can be performed separately which gives two scaling functions for the infinite system.
Additionally it is very important to point out that $\tilde{\Lambda}$ is always decaying with system size. Usually this is 
interpreted as localizing behavior. Whereas our analysis shows a rather unusual phase transition, namely that the correlation 
length diverges only when approaching the critical point from below $W_c$. In order to extract the functional behavior of 
$\xi$ at the transition point we fitted the data to several functions and found best agreement with
\begin{equation}
 \xi(W) \propto 
|W-W_c|^{-\nu_L} \ \ \  \text{for } 0< W <W_c
\, .
\label{eq:scaling-function-gap02-delta0}
\end{equation}
The results for the critical parameters are 
\[
\nu\approx 0.289\pm 0.013 \ \ \ (W_c\approx 2.156\pm 0.009) 
\ .
\]
If we compare the variance $g$ of the 
fitted critical disorder strength which is $g_c=0.387$ to the gap width $2\bar{m}=0.4$ we see a good agreement. 
A possible explanation for this might be that if fluctuations of the random gap are larger than the gap width 
states are no more exponentially localized and diffusive transport is possible. From this point of view we can 
also calculate $W_c$ from the average gap width which yields $\tilde{W}_c 
\approx 2.191$. 
Fitting~(\ref{eq:scaling-function-gap02-delta0}) with fixed critical disorder gives slightly different exponents 
but also a very good agreement with the numerical scaling function for $0<W<W_c$:
\[
\nu\approx 0.332\pm 0.004 \ \ \ (W_c\approx 2.191)
\ .
\]





\subsection{Broken node symmetry: $\delta\ne0$}

\begin{figure}[ht]
\centering
 \subfigure{
 \includegraphics[scale=0.65]{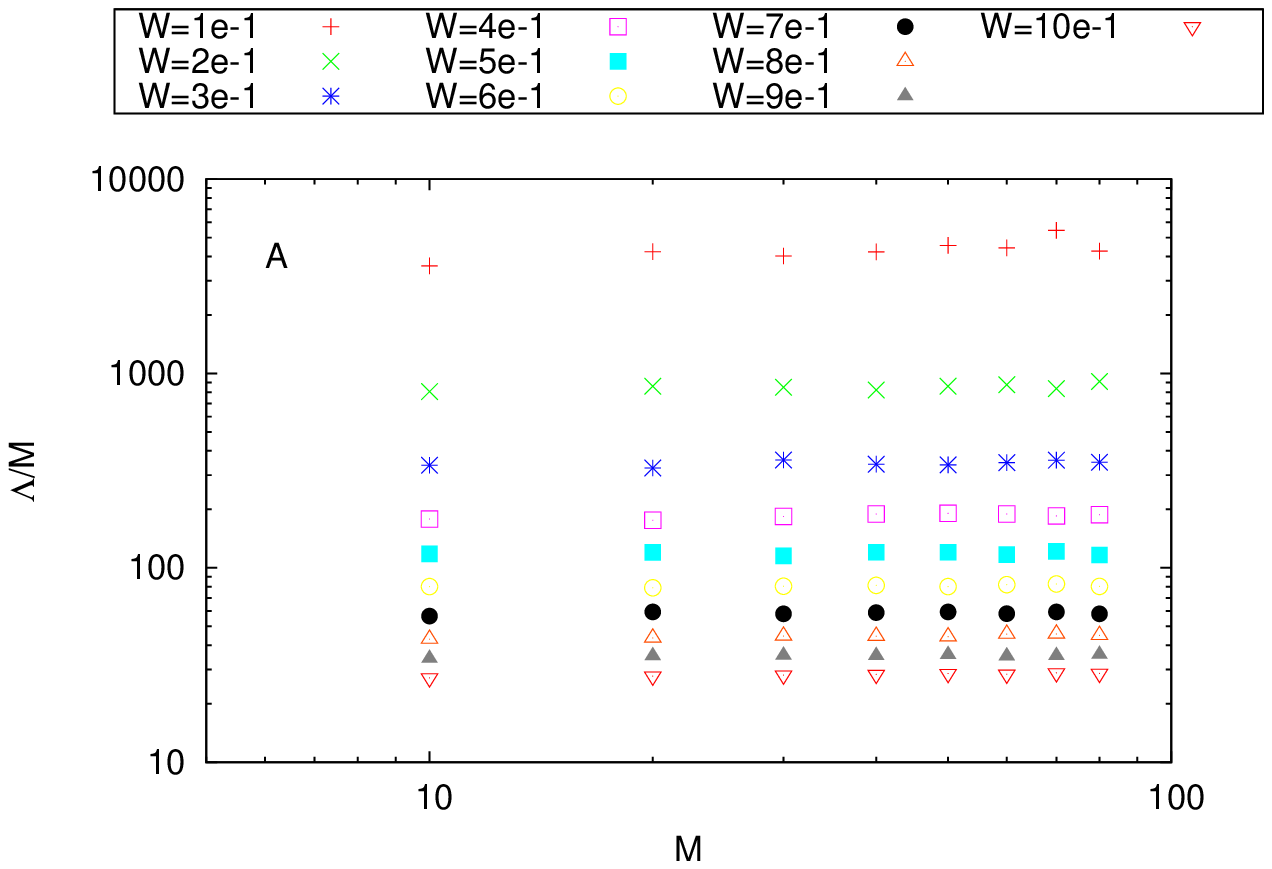}
\label{fig:loc-len-gap0-delta05-small-disorder_a}
 }
\subfigure{
   \includegraphics[scale=0.65]{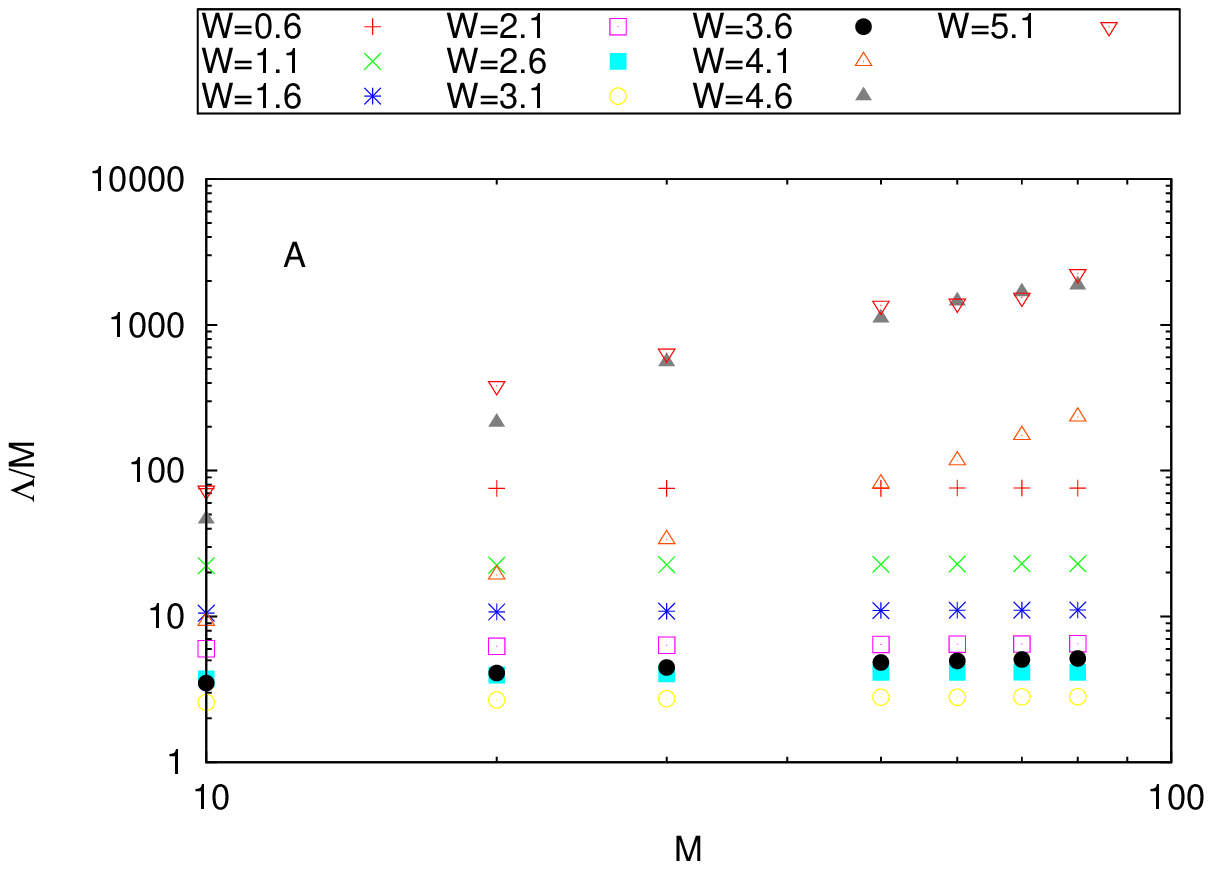}
\label{fig:loc-len-gap0-delta05_a}
 }
\caption{Localization length for systems with zero average gap $\bar{m}=0$ and broken node symmetry $\delta =0.5$. Left and right panel are for different disorder ranges.}
\label{fig:loc-len-gap0-delta05-compare}
\end{figure}

By setting $\delta=0.5$ we break the four-fold degeneracy of the node structure and retain only the node at $k_x=k_y=0$.
Unlike in the case of $\delta=0$ the localization length is not growing with system size if $\bar{m}=0$. 
Fig.~\ref{fig:loc-len-gap0-delta05-small-disorder_a} shows that for weak disorder $\Lambda$ is constant with 
increasing $M$ but decreases with increasing disorder $W$. However, for  $W\geq4.1$ $\Lambda$ it increases with $M$ (Fig.~\ref{fig:loc-len-gap0-delta05_a}). 
The normalized data is shown in Fig.~\ref{fig:loc-len-over-disorder-gap0-delta05_a}.
To keep the plot illustrative only a choice of the whole data is shown. 
What can be seen in Fig.~\ref{fig:loc-len-over-disorder-gap0-delta05_a}
is that for weak disorder up to $W=3.6$ the normalized localization 
length decays for growing system sizes and scales to zero with $M$. For disorder larger than $W=3.6$ $\tilde{\Lambda}$ is growing with system size.
The growing localization length may be explained by comparison to the clean case. If fluctuations of the random gap are in the range of $2\delta$ a massless fermion appears. Thus disorder effectively closes the gap at the border of the Brillouin zone and the model shows metallic behavior.

\begin{figure}[ht]
\centering
\subfigure{
 \includegraphics[scale=0.55]{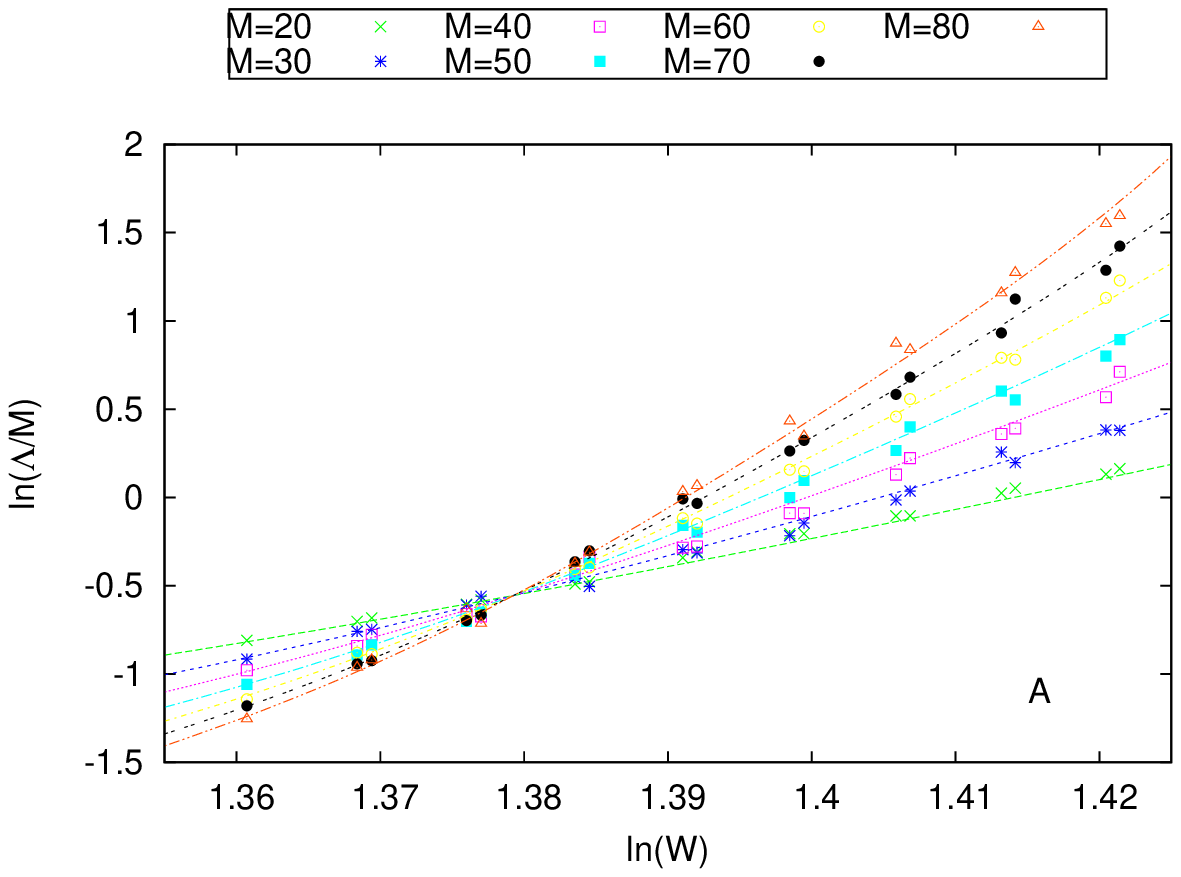}  
 }
 \subfigure{
\includegraphics[scale=0.55]{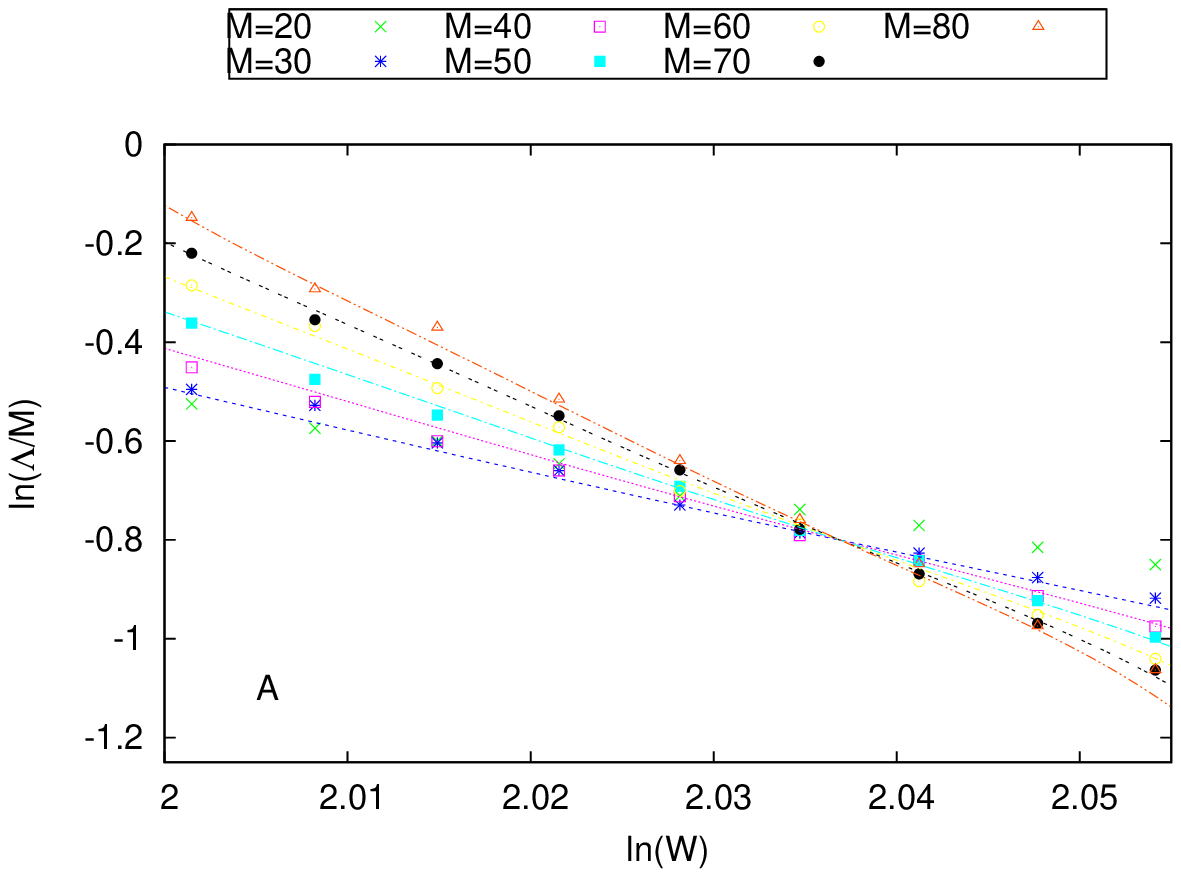}
 }
\caption{Fits to Eq.~(\ref{eq:fit-exponent}) for $\bar{m}= 0$ and $\delta=0.5$ around the critical point I (left) and around the critical point II (right).}
\label{fig:fitted-curves-gap0-delta05_a}
\end{figure}

For weak (i.e. $W\lessapprox 4$) and strong disorder (i.e. $W\gtrapprox 7.5$) the behavior is qualitatively the same,
characterized by a decaying behavior of $\tilde{\Lambda}$ with increasing $M$. The benefit of plotting $\tilde{\Lambda}$ 
over $W$ is that one can see directly two scale invariant points where different $\tilde{\Lambda}$ are intersecting 
for all available values of $M$. These points are indicative of phase transitions. Now we use the fitting functions of
Eq. (\ref{eq:fit-exponent}) to extract the critical exponent $\nu$ from our numerical result. For this purpose we set
$s=5$ and obtain the resulting curves in Fig.~\ref{fig:fitted-curves-gap0-delta05_a}.
The critical parameters are listed in table~\ref{table:critical-values-gap0-delta05}. 

\begin{figure}[ht]
\centering
\subfigure{
 \includegraphics[scale=0.55]{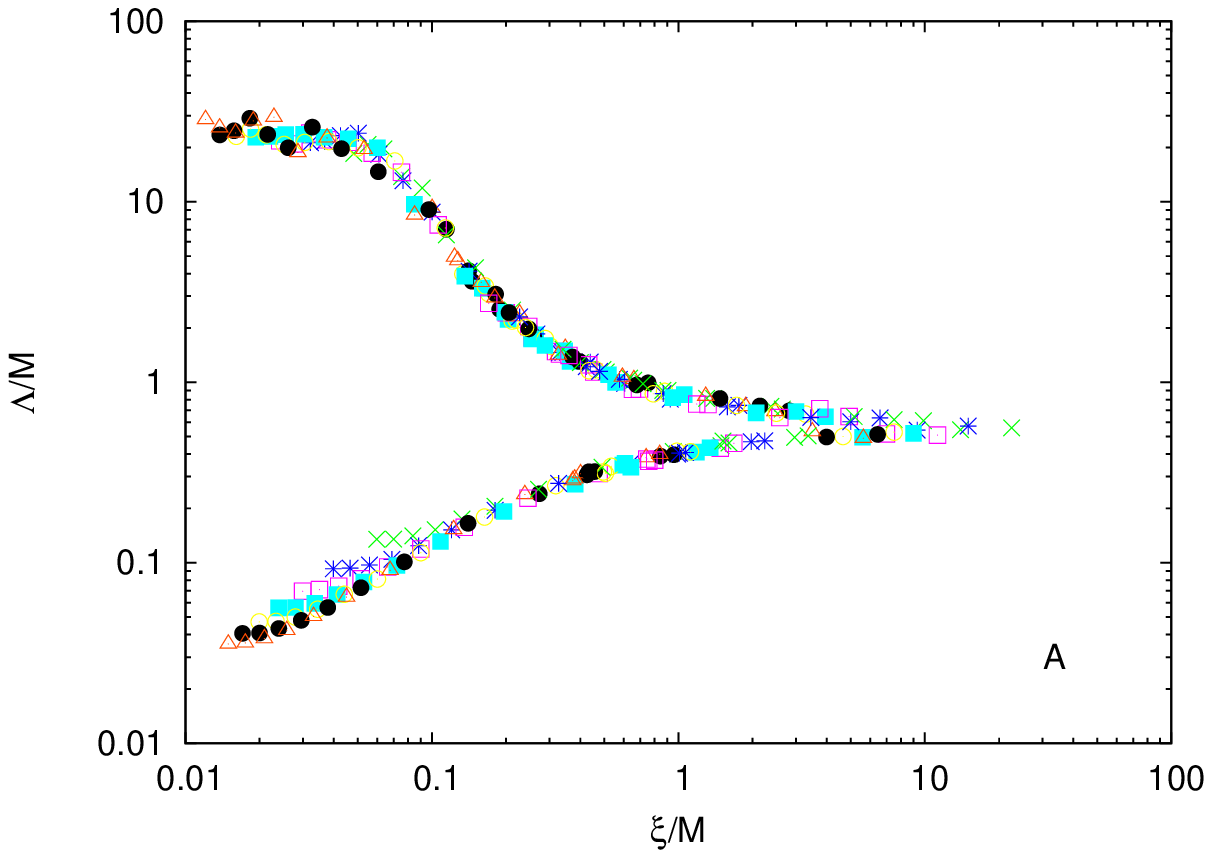}  
 }
 \subfigure{
\includegraphics[scale=0.55]{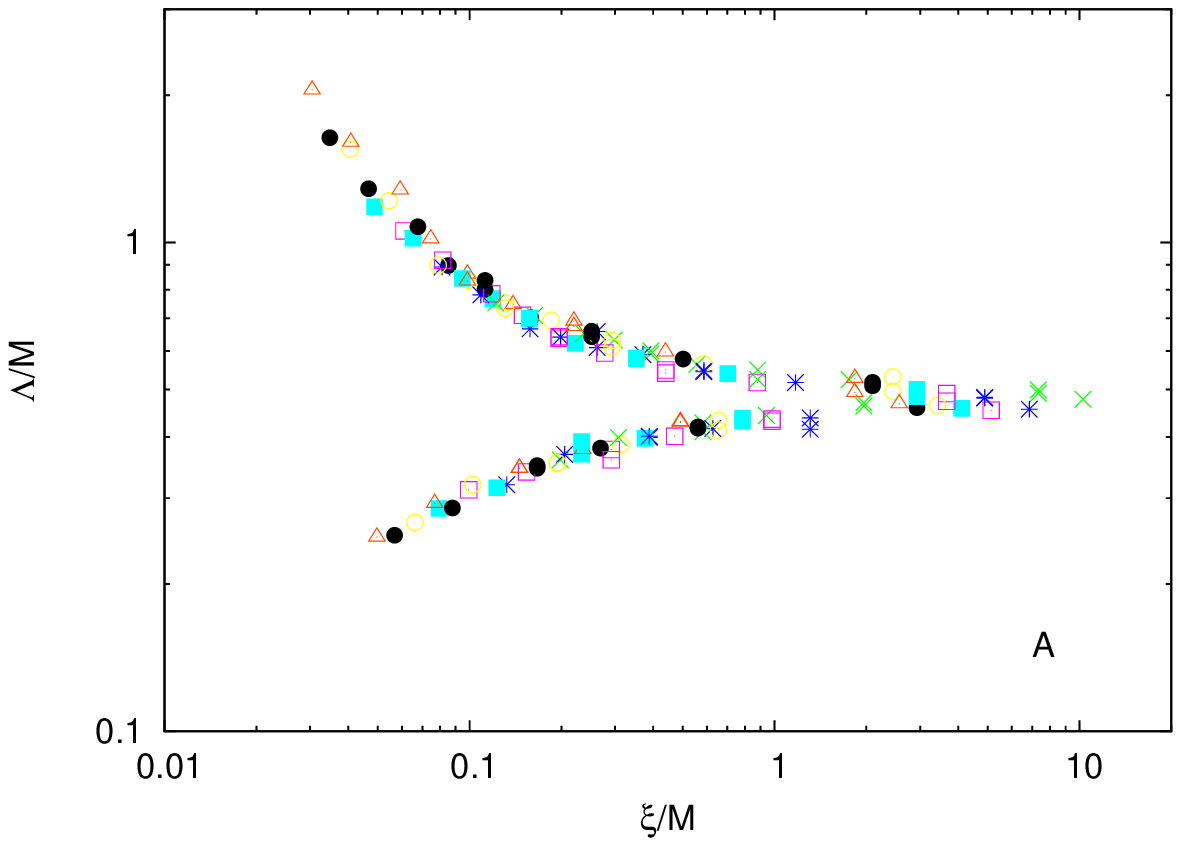}
 }
\caption{Rescaled NLL for $\bar{m}= 0$ and $\delta=0.5$ around the critical point I (left) and around the critical point II (right). 
Plots contain more data points than used for the fitting procedure to show the validity of one parameter scaling.}
\label{fig:rescaled-nll-gap0-delta05_a}
\end{figure}

\begin{figure}[ht]
\centering
\subfigure{
 \includegraphics[scale=0.55]{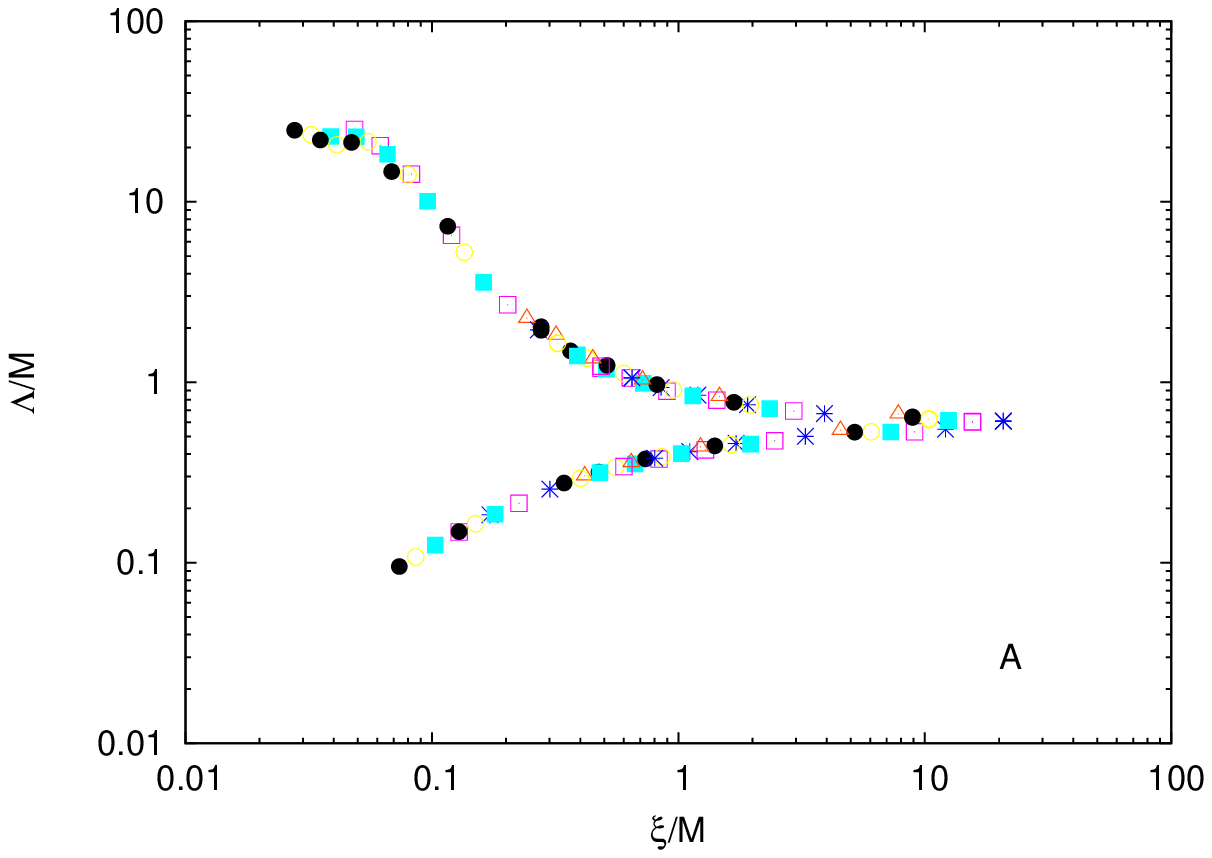}  
 }
 \subfigure{
\includegraphics[scale=0.55]{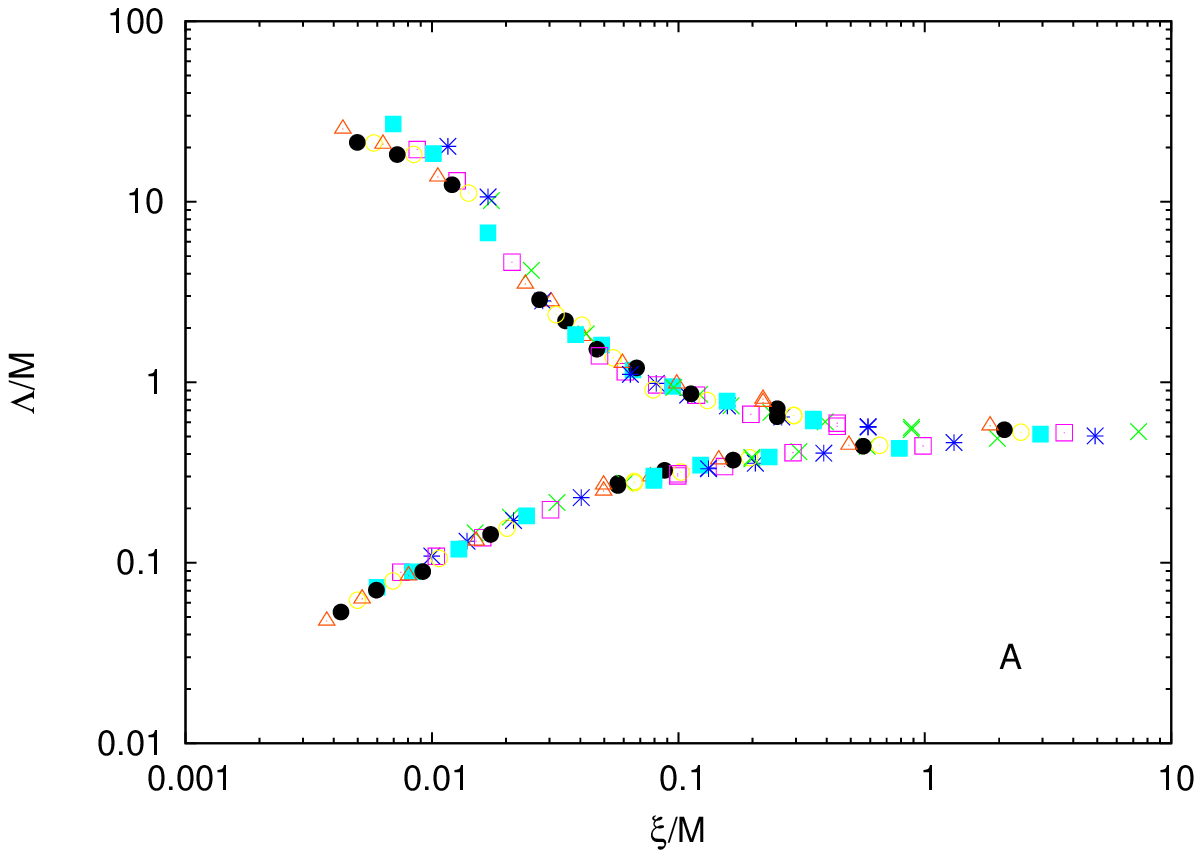}
 }
\caption{Rescaled NLL for $\bar{m}= 0.2$ and $\delta=0.5$ in the vicinity of critical point I (left) and in the vicinity of critical point II (right).}
\label{fig:rescaled-nll-gap02-delta05_a}
\end{figure}

Using the scaling form of Eq. (\ref{eq:scaling-sol}) all the curves collapse on two curves for a proper choice of the scaling
function $\xi$, as depicted in Figs.~\ref{fig:rescaled-nll-gap0-delta05_a}, \ref{fig:rescaled-nll-gap02-delta05_a}. 
There plots agree qualitatively well for ${\bar m}=0$ and ${\bar m}=0.2$, the critical exponents for the second transition 
differ slightly though (cf. tables~\ref{table:critical-values-gap0-delta05} and \ref{table:critical-values-gap02-delta05}).



\begin{table}[ht]
\centering
    \begin{tabular}{ l  c  c  }
    Critical point & I & II \\ \hline\hline
	Exponent $\nu$ & $1.297\pm 0.031$ & $1.299\pm 0.066$  \\ 
	$W_c$ & $3.975\pm 0.002$ & $7.668\pm0.008$  \\
	$\Lambda_c$ & $0.574\pm 0.009$ & $0.447\pm0.005$\\
	Disorder range & $3.87 \leq W \leq 4.17$ & $7.35\leq W \leq 7.8$ \\ 
	System sizes & $20 \leq M \leq 80$ & $30 \leq M \leq 80$ \\
	\hline
    \end{tabular}
    \caption{Critical values for $\bar{m}= 0$ and $\delta=0.5$ obtained from fitting the data to equation~(\ref{eq:fit-exponent}).}
    \label{table:critical-values-gap0-delta05}
\end{table}
\begin{table}[ht]
\centering
    \begin{tabular}{ l  c  c  }
    Critical point & I & II \\ \hline\hline
	Exponent $\nu$ & $1.297\pm 0.045$ & $1.397 \pm 0.069$  \\ 
	$W_c$ & $3.792\pm 0.002$ & $7.629 \pm 0.015$  \\
	$\Lambda_c$ & $0.591\pm 0.007$ & $0.517 \pm 0.009$\\
	Disorder range & $3.72 \leq W \leq 3.88$ & $7.1\leq W \leq 8.0$ \\ 
	System sizes & $20 \leq M \leq 80$ & $20 \leq M \leq 80$ \\
	\hline
    \end{tabular}
    \caption{Critical values for $\bar{m}=0.2$ and $\delta=0.5$ obtained from fitting the data to equation~(\ref{eq:fit-exponent}).}
    \label{table:critical-values-gap02-delta05}
\end{table}

\section{Discussion}

Our numerical results can be summarized as follows. The localization length $\Lambda$ always increase with
$M$ according to the power law of Eq. (\ref{eq:LL-power-law}), where the exponent $\alpha$ depends on the
model parameters:
\begin{equation}
\begin{cases}
0<\alpha<1 & \text{for}\ \  \delta=0, {\bar m}=0 \\
\alpha=0 & \text{for}\ \  \delta=0, {\bar m}=0.2, W \le W_c\\
0<\alpha<1 & \text{for}\ \  \delta=0, {\bar m}=0.2, W > W_c\\
\alpha=0 & \text{for}\ \  \delta=0.5, {\bar m}=0, 0.2, W \le W_{c1}\\
\alpha>1 & \text{for}\ \  \delta=0.5, {\bar m}=0, 0.2, W_{c1} \le W\le W_{c2}\\
0<\alpha<1 & \text{for}\ \  \delta=0.5, {\bar m}=0, 0.2, W > W_{c2}
\end{cases}
\ ,
\label{exponents}
\end{equation}
where $\delta =0$ represents the case with four degenerate nodes and $\delta=0.5$ a single node.
In our numerical results we can distinguish these to  two cases: 
(I) For a preserved four-fold node degeneracy (i.e. $\delta=0$)
the gapless system has a monotonically increasing localization length with $M$ as well as with $W$ and
does not indicate any transition. In the presence of a gap (${\bar m}\ne 0$), however, there is a qualitative change
at a characteristic disorder strength $W_c$: For $W<W_c$ the states are exponentially localized, whereas
for $W>W_c$ they are not. It is not possible to decide within our numerical approach whether they are really 
extended or power-law localized in the gapped case.  As discussed in Appendix A, it might be sufficient for diffusion
in a 2D system that the states obey a power law.

(II) For the single node (i.e. $\delta=0.5$) the one-parameter scaling analysis of our results indicates a 
typical Anderson transition at two critical points $W_{c1}$, $W_{c2}$. 
The exponent $\alpha=0$ for weak disorder (i.e. for $W<W_{c1}$) indicates exponentially localized states.
There is the intermediate metallic phase for $W_{c1}<W<W_{c2}$ with $\alpha>1$
with one-parameter scaling behavior near the critical points. This is indicative of two metal-insulator transitions.
In particular, there is a metal-insulator transition from $\alpha=0$ to $\alpha>1$ at a critical $W_{c1}$, which
corresponds to a transition from $\alpha=0$ to $0<\alpha<1$ for the gapped four degenerate Dirac nodes. The difference
between a transition from $\alpha=0$ to $0<\alpha<1$ and  a transition from $\alpha=0$ to $\alpha>1$ is not
clear from our numerical results. It could be that the latter is a genuine transition from exponentially localized to
extended states, whereas the former is a transition from exponentially localized to power-law localized states.

\section{Conclusion}

We have introduced a model for Dirac fermions on a lattice with several nodes which allows us to perform numerical 
calculations of the localization length within the frame work of the transfer matrix formalism. 
Using the Hamiltonian in Eq. (\ref{main_ham})
it is possible to break the node symmetry and to compare the properties for one and four nodal points in the Brillouin zone.
We have shown that states in the gap can be localized and thus the localization length $\Lambda$ converges to a finite value 
for increasing system size, whereas in the gapless case there are extended states as expected. 

We have calculated the localization length for various system sizes and for different strength of the random gap.
In all cases the localization length grows like a power law $\Lambda\sim M^\alpha$ with increasing system width $M$.
However, the exponent $\alpha$ is quite sensitive to the model parameters (cf. (\ref{exponents})). In particular, this
exponent vanishes for nonzero average gap and weak disorder, indicating exponential localization. Our numerical
result also indicates $\alpha=0$ for non-degenerate nodes, vanishing gap and weak disorder. On the other hand,
we have $\alpha>1$ only for intermediate disorder strength and non-degenerate nodes. Thus, the nodal degeneracy
 suppresses the intermediate phase. The latter is separated from the phases with $0\le \alpha<1$ by transitions that
obey one-parameter scaling behavior with scale-invariant critical points. This reflects the results
of the weak-localization theory, where (anti-)localization has been found for (single) two nodes 
\cite{ando98,ando02}.

\appendix

\section{Localization and Diffusion in 2D}

Exponentially localized states rule out diffusive behavior. Here we briefly discuss that
a power-law decaying state can provide diffusive behavior in a 2D electron gas.  
Diffusion of $|\Psi(\br,t)|^2$ in 2D is defined by the diffusion equation
\begin{equation}
\left(\frac{\partial}{\partial t}-\frac{D}{4}\nabla^2\right)|\Psi(\br,t)|^2=0
\ ,
\label{diff_eq}
\end{equation}
which has an expanding solution
\[
|\Psi(\br,t)|^2\equiv K(\br, \omega)
\sim\frac{e^{-r^2/Dt}}{Dt} \ \ \ (t\sim\infty) 
\ .
\]
The solution of Eq. (\ref{diff_eq}) is also given by the diffusion propagator
\[
{\tilde K}(q,\omega)=\frac{\bar K}{\omega+Dq^2}
\ .
\]
On the other hand, the localization length $\xi$ in the spatial direction $j$ can be defined as
\[ 
\xi= 
\sqrt{\sum_\br r_j^2 K(\br, \omega)}
\ ,
\]
where $K(\br, \omega)$ is connected with the diffusion propagator by a Fourier transformation:
\[
K(\br,\omega)={\bar K}\int \frac{e^{i\bq\cdot\br}}{\omega+Dq^2}d^2q
\ .
\]
Using the Bessel function $J_0$ and the momentum cut-off $\lambda$ for the $q$ integral this result leads to
\[
K(\br,\omega=0)-K(\br',\omega=0)=\frac{{\bar K}}{D}\int_{\lambda r'}^{\lambda r}\frac{J_0(x)}{x}dx
\]
and for $\lambda r',\lambda r\gg 1$
\[
\sim \frac{{\bar K}}{D}\sqrt{\frac{2}{\pi}}
\int_{\lambda r'}^{\lambda r}\frac{\cos(x-\pi/4)}{x^{3/2}}dx
\ .
\]
Thus $K(\br,\omega=0)$ decays on large scales like $r^{-1/2}$. This reflects the fact that a decaying
wave function leads to diffusion in 2D.


\begin{thebibliography}{99}

\bibitem{novoselov05}
K. S. Novoselov, A. K. Geim, S. V. Morozov, D. Jiang, M. I. Katsnelson, I. V. Grigorieva, S. V. Dubonos,
A. A. Firsov, Nature {\bf 438}, 197 (2005).

\bibitem{zhang05}
Y. Zhang, Y.-W. Tan, H. L. Stormer, P. Kim, Nature {\bf 438}, 201 (2005).

\bibitem{ando98}
N. H. Shon and T. Ando, J. Phys. Soc. Japan {\bf 67}, 2421 (1998).

\bibitem{beenakker08}
J. Tworzyd\l o, C.W. Groth and C.W.J. Beenakker, Phys. Rev. B {\bf 78}, 235438 (2008).

\bibitem{ando02}
H. Suzuura and T. Ando, Phys. Rev. Lett. {\bf 89}, 266603 (2002). 

\bibitem{MacKinnon1983}
A. MacKinnon and B. Kramer, Z. Phys. B Condensed Matter {\bf 13}, 1546 (1983).

\bibitem{Pichard1981}
J. L. Pichard and G. Sarma, J. Phys. C: Solid State Phys. {\bf 14}, L127 (1981).

\bibitem{bostwick09}
A. Bostwick, J. L. McChesney, K.V. Emtsev, Th. Seyller, K. Horn, S. D. Kevan, and E. Rotenberg,
Phys. Rev. Lett. {\bf 103}, 056404 (2009).

\bibitem{elias2009}
D. C. Elias, R. R. Nair, T. M. G. Mohiuddin, S. V. Morozov, P. Blake, 
M. P. Halsall, A. C. Ferrari, D. W. Boukhvalov, M. I. Katsnelson, 
A. K. Geim and K. S. Novoselov, Science {\bf 323}, 610 (2009).

\bibitem{geim11}
L. A. Ponomarenko, A. K. Geim, A. A. Zhukov, R. Jalil, S. V. Morozov, K. S. Novoselov, V. V. Cheianov, V. I. Fal'ko, K. Watanabe, T. Taniguchi and R. V. Gorbachev, Nature Physics {\bf 7}, 958 (2011).

\bibitem{abergel10}
D. S. L. Abergel, V. Apalkov, J. Berashevich, K. Ziegler and T. Chakraborty, Adv. Phys. {\bf 59}, 261  (2010).

\bibitem{Ziegler2009} 
K. Ziegler, Phys. Rev. Lett. {\bf 102}, 126802 (2009); Phys. Rev. B {\bf 79}, 195424 (2009).

\bibitem{Susskind1977}
L. Susskind, Phys. Rev. D {\bf 16}, 3031 (1977).

\bibitem{Stacey1982}
R. Stacey, Phys. Rev. D {\bf 26}, 468 (1982).

\bibitem{beenakker10}
M. V. Medvedyeva, J. Tworzyd\l o, and C. W. J. Beenakker, Phys. Rev. B {\bf 81}, 214203 (2010).

\bibitem{Ziegler1996}
K. Ziegler, Phys. Rev. B {\bf 53}, 9653 (1996).

\bibitem{oseledec}
V. Oseledec, Trans. Moscow Math. Soc. {\bf 19}, 197 (1968).

\bibitem{MacKinnon1981}
A. MacKinnon and B. Kramer, Phys. Rev. Lett. {\bf 47}, 21 (1981).











































\end{thebibliography}
\end{document}